# On the Use of Suffix Arrays for Memory-Efficient Lempel-Ziv Data Compression

Artur J. Ferreira[1,3], Arlindo L. Oliveira[2,4], Mário A. T. Figueiredo[3,4]

[1]*Instituto Superior de Engenharia de Lisboa, Lisboa, PORTUGAL*
[2]*Instituto de Engenharia de Sistemas e Computadores, Lisboa, PORTUGAL*
[3]*Instituto de Telecomunicações, Lisboa, PORTUGAL*
[4]*Instituto Superior Técnico, Lisboa, PORTUGAL*
Contact email: `arturj@cc.isel.ipl.pt`

**Abstract**

Much research has been devoted to optimizing algorithms of the Lempel-Ziv (LZ) 77 family, both in terms of speed and memory requirements. Binary search trees and *suffix trees* (ST) are data structures that have been often used for this purpose, as they allow fast searches at the expense of memory usage.

In recent years, there has been interest on *suffix arrays* (SA), due to their simplicity and low memory requirements. One key issue is that an SA can solve the sub-string problem almost as efficiently as an ST, using less memory. This paper proposes two new SA-based algorithms for LZ encoding, which require no modifications on the decoder side. Experimental results on standard benchmarks show that our algorithms, though not faster, use 3 to 5 times less memory than the ST counterparts. Another important feature of our SA-based algorithms is that the amount of memory is independent of the text to search, thus the memory that has to be allocated can be defined *a priori*. These features of low and predictable memory requirements are of the utmost importance in several scenarios, such as embedded systems, where memory is at a premium and speed is not critical. Finally, we point out that the new algorithms are general, in the sense that they are adequate for applications other than LZ compression, such as text retrieval and forward/backward sub-string search.

## 1 Introduction

It is well known that LZ coding is very asymmetric in terms of time and memory requirements, with encoding being much more demanding than decoding [1, 2, 3]. A significant amount of research has been devoted to optimizing LZ encoding, namely by devising efficient data structures, like *suffix trees* (ST) [4, 5], which have been deeply explored [6].

Recently, attention has been payed to *suffix arrays* (SA) [4, 7], due to their simplicity and space efficiency; moreover, linear-time SA construction algorithms are known [8, 9]. SA have been successfully used in search, indexing, plagiarism detection, information retrieval, biological sequence analysis [10]. In data compression, SA have been used in encoding data with anti-dictionaries [11] and optimized for large alphabets [12]. The space requirement issue of ST has been addressed by replacing ST-based algorithms with methods based on (*enhanced*) SA [13].

In [14], we have proposed a method in which an SA replaces an ST, to hold the dictionary in the LZ77 and LZSS algorithms [1, 3]. We have shown that an SA-based encoder requires less memory than an ST-based encoder, with a small penalty on the encoding time, for roughly the same compression ratio. We also have shown that the amount of memory for the SA-based encoder is constant, independent of the contents of the sequence to encode; this may not be true for an ST or tree-based encoder. In this paper, we further explore the approach proposed in [14], presenting two new encoding algorithms for SA-based LZ77/LZSS encoders. We focus only on the encoder data structures and algorithms, since no modifications are required on the corresponding decoders.

The rest of the paper is organized as follows. Section 2 contains brief reviews of the LZ77 and LZSS algorithms, as well as of the main features of SA. In Section 3, we describe the proposed algorithms, while Section 4 reports experimental results on standard benchmarks. Finally, some concluding remarks are made in Section 5.

## 2 Background

### 2.1 The LZ77 and LZSS Algorithms

The well-known LZ77 and LZSS encoding algorithms use a sliding window over the sequence of symbols, which has two sub-windows: the *dictionary* (holding the symbols already encoded) and the *look-ahead-buffer* (LAB, containing the symbols still to be encoded) [1, 2]. As a string of symbols in the LAB is encoded, the window slides to include it in the dictionary (this string is said to *slide in*); consequently, symbols at the far end of the dictionary are dropped (they *slide out*).

At each step of the LZ77/LZSS encoding algorithm, the longest prefix of the LAB which can be found anywhere in the dictionary is determined and its position stored. In the example of Fig. 1, the string of the first four LAB symbols ("brow") is found in position 17 of the dictionary. For these two algorithms, encoding of a string consists in describing it by a token. The LZ77 token is a triplet of fields (*pos, len, symbol*), with the following meanings:

- *pos* - location of the longest prefix of the LAB found in the current dictionary; this field uses $\log_2(|\text{dict.}|)$ bits, where |dict.| is the length of the dictionary;

- *len* - length of the matched string; this requires $\log_2(|\text{LAB}|)$ bits;

- *symbol* - the first symbol in the LAB, that does not belong to the matched string (*i.e.*, that breaks the match); for ASCII symbols, this uses 8 bits.

In the absence of a match, the LZ77 token is (*0,0,symbol*).

In LZSS, the token has the format (*bit,code*), with the structure of *code* depending on value *bit* as follows:

$$\begin{cases} bit = 0 & \Rightarrow \quad code = (symbol), \\ bit = 1 & \Rightarrow \quad code = (pos,\ len). \end{cases} \quad (1)$$

In the absence of a match, LZSS produces (0(*symbol*)). The idea is that, when a match exists, there is no need to explicitly encode the next symbol. Besides this modification, Storer and Szymanski [3] also proposed keeping the LAB in a circular queue and the dictionary in a binary search tree, to optimize the search. LZSS is widely used in practice (*e.g.*, in GZIP and PKZIP) since it typically achieves higher compression ratios than LZ77.

Fig. 1 illustrates LZ77 and LZSS encoding. In LZ77, the string "brows" is encoded by (17,4,s); the window then slides 5 positions forward, thus the string "after" *slides out*, while the string "brows" *slides in*. In LZSS, "brow" is encoded as (1(17,4)) and "brow" *slides in*. Each LZ77 token uses $\log_2(|\text{dict.}|) + \log_2(|\text{LAB}|) + 8$ bits; usually, $|\text{dict.}| \gg |\text{LAB}|$. In LZSS, the token uses either 9 bits, when it has the form (0,(symbol)), or $1 + \log_2(|\text{dict.}|) + \log_2(|\text{LAB}|)$ bits, when it has the form (1,(position,length)).

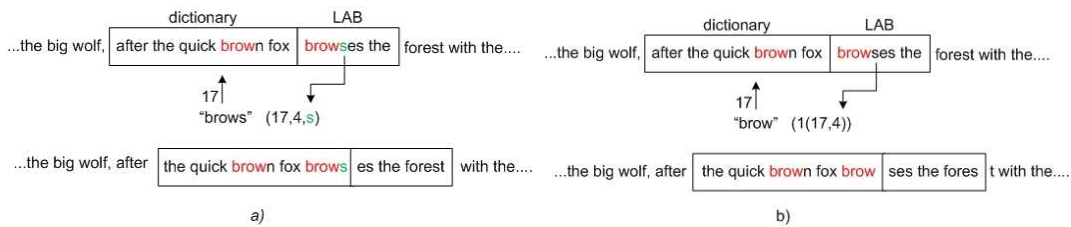

Figure 1: a) LZ77 encoding of string "brows", with token (17,4,'s') b) LZSS encoding of string "brow", with token (1(17,4)).

It is clear that the key component of the LZ77/LZSS encoding algorithms is the search for the longest match between LAB prefixes and the dictionary. Recently, ST have been used as efficient data structures to support this search [6].

### 2.1.1 Decoding

Assuming the decoder and encoder are initialized with equal dictionaries, the decoding of each LZ77 token (*pos,len,symbol*) proceeds as follows: (1) *len* symbols are copied from the dictionary to the output, starting at position *pos* of the dictionary; (2) the symbol *symbol* is appended to the output; (3) the string just produced at the output is slid into the dictionary.

For LZSS, we have: if the bit field is 1, *len* symbols, starting at position *pos* of the dictionary, are copied to the output; if it is 0, *symbol* is copied to the output; finally, the string just produced at the output is slid into the dictionary.

Clearly, both LZ77/LZSS decoding are low complexity procedures. In this work, we address only encoder data structures and algorithms, with no effect in the decoder.

## 2.2 Suffix Arrays

A suffix array (SA) is the lexicographically sorted array of the suffixes of a string [4, 7]. For a string $D$ of length $m$ (with $m$ suffixes), an SA $P$ is a list of integers from 1 to $m$, according to the lexicographic order of the suffixes of $D$. For instance, if we consider $D = mississippi$ (with $m = 11$), we get the suffixes in Fig. 2 part a); after sorting, we get the suffixes in part b). Thus, the SA for $D$ is $P = \{11, 8, 5, 2, 1, 10, 9, 7, 4, 6, 3\}$. Each of these integers is the suffix number, therefore corresponding to its position in $D$. Finding a sub-string of $D$ can be done by searching vector $P$; for instance, the set of sub-strings of $D$ that start with symbol 's', can be found at positions 7, 4, 6, and 3. As a result of this, an SA can be used to obtain every occurrence of a sub-string within a given string. For LZ77/LZSS encoding, we can find the set of sub-strings of $D$, starting with a given symbol (the first in the LAB).

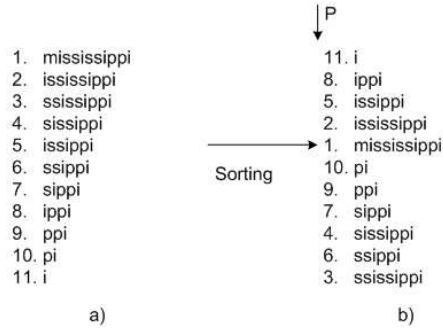

Figure 2: Suffixes of $D = mississippi$: a) Set of suffixes b) Sorted suffixes and SA $P$.

SA are an alternative to ST, as an SA implicitly holds the same information as an ST. Typically, it requires $3 \sim 5$ times less memory and can be used to solve the sub-string problem almost as efficiently as an ST [4]; moreover, its use is more appropriate when the alphabet is large. An SA can be built using a sorting algorithm, such as "quicksort"; it is possible to convert an ST into an SA in linear time [4]. An SA does *not* has the limitation of an ST: no suffix of smaller length prefixes another suffix of greater length.

## 2.3 Longest Common Prefix

The lexicographic order of the suffixes implies that suffixes that start with the same symbol are consecutive on SA $P$. This means that a binary search on $P$ can be used to find all these suffixes; this search takes $\mathcal{O}(n \log(m))$ time, with $n$ being the length of the sub-string to find, while $m$ is the length of the dictionary. To avoid some redundant comparisons on this binary search, the use of *longest common prefix* (LCP) of the suffixes, lowers the search time to $\mathcal{O}(n + \log(m))$ [4]; the computation of LCP takes $\mathcal{O}(m)$ time.

The LCP of two suffixes $i$ and $j$, named $Lcp(i,j)$, is defined as the length of the longest common prefix of the suffixes located at positions $i$ and $j$ of $P$. An alternative definition is $Lcp(i,j)$ **is the length of the longest prefix common to suffix** $P(i)$ **and** $P(j)$. For the SA $P$, of Section 2.2 and Fig. 2, we get the LCP array $L = \{0, 1, 1, 4, 0, 0, 1, 0, 2, 1, 3\}$, as depicted in Fig. 3. For instance, suffix 2 has 4 symbols in common with suffix 5.

## 3 LZ77/LZSS Compression Using Suffix Arrays

This section presents three algorithms for LZ77/LZSS encoding. The first is a minor modification of the algorithm proposed in [14], while the other two are new proposals. We stress

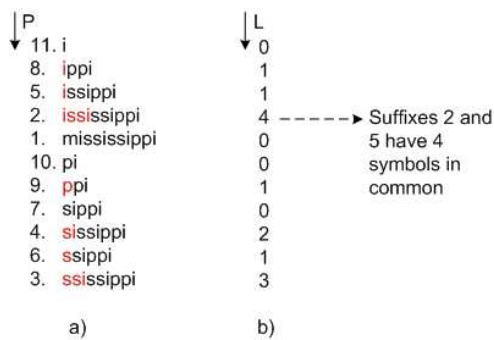

Figure 3: SA for $D = mississippi$ (part a)) and corresponding LCP values $L$ (part b)).

that these are fast sub-string search algorithms, thus are not limited to LZ encoding.

## 3.1 Algorithm 1

SA-based LZ77/LZSS encoding can be carried out using Algorithm 1, in which an SA $P$ is built for each dictionary. To encode the sequence in the LAB, we search the SA $P$ for the (longest) sub-string that starts with the same symbol as the prefix of the LAB. After encoding the contents of the LAB, these symbols are slid into the dictionary (and the corresponding slide out is performed). Next, a new SA for the new dictionary is built, and the procedure is repeated until the end of stream is reached. The encoder data structures are the dictionary symbols (string) and the SA $P$, that is, an $m$-length string and an $m$-length integer array.

---
**A1 - Algorithm 1 - Simple SA-Based LZSS Encoding**

Inputs: $In$, input stream to encode; $m$ and $n$ length of dictionary and LAB.
Output: $Out$, output stream with LZSS description of $In$.

---

1. Read dictionary $D$ and look-ahead-buffer $LAB$ from $In$;
2. While there are symbols of $In$ to encode:
   a) Build SA for string $D$ and name it $P$;
   
   b) To obtain the description ($pos,len$) for every sub-string $LAB[i \ldots n]$, $1 \leq i \leq n$, proceed as follows:
   
   b1) Do a binary search on vector $P$ until we find:
   
   i) the first position $left$, in which the first symbol of the corresponding suffix matches $S[i]$, that is, $D[P[left]] = S[i]$;
   
   ii) the last position $right$, in which the first symbol of the corresponding suffix matches $S[i]$, that is, $D[P[right]] = S[i]$;
   
   If no suffix starts with $S[i]$, output $(0(S[i]))$ to $Out$, set $i \leftarrow i + 1$ and goto 2b).
   
   b2) From the set of suffixes between $P[left]$ and $P[right]$, choose the $k^{th}$ suffix, $left \leq k \leq right$, with a given criteria (see below) giving a p-length match.
   
   b3) Do $pos \leftarrow k$ and $len \leftarrow p$ and output token $(1(pos, len))$ into $Out$.
   
   b4) Do $i \leftarrow i + len$; if $i = n$ stop; else goto 2b).
   
   c) Slide in the full encoded LAB into $D$;

d) Read next LAB from *In*; goto 2).

---

In step 2.b2), it is possible to choose among several suffixes, according to a greedy/non-greedy parsing criterion. If we seek a fast search, we can choose one of the immediate suffixes, given by left or right. If we want better compression ratio, at the expense of a not so fast search, we should choose the suffix with the longest match with sub-string $LAB[i \ldots n]$.

## 3.2 Algorithm 2

The second algorithm does not build a new SA every time the LAB is encoded. Instead, we build a single SA for the initial dictionary and every time we reach the end of the LAB, we just slide the LAB into the dictionary and update the indexes of the SA (built at the beginning of the encoding process). This way, we have a sliding window suffix array. As in Algorithm 1, the encoder data structures are composed solely by the dictionary symbols (string) and the SA $P$ (an $m$-length string and an $m$-length integer array).

---

**A2 - Algorithm 2 - SA-Based LZSS Encoding**

Inputs: $In$, input stream to encode; $m$ and $n$ length of dictionary and LAB.
Output: $Out$, output stream with LZSS description of $In$.

1. Read dictionary $D$ and look-ahead-buffer $LAB$ from $In$;
2. **Build SA for string** $D$ and name it $P$;
3. While there are symbols of $In$ to encode:
    a) To obtain the description (*pos,len*) for every sub-string $LAB[i \ldots n]$, $1 \leq i \leq n$, proceed as in Algorithm 1;
    b) Slide in LAB into $D$;
    c) Read next LAB from $In$;
    d) **Update $P$ using the SA for LAB (see below)**;

---

The ideas of the update (step 3d) are: after each full LAB encoding and corresponding slide in, the resulting dictionary is (very closely) related to the previous one; it is faster to build an SA for the LAB instead of an SA for the dictionary. After we encode _the, we can see that the resulting dictionary is obtained by the following actions (step 3d)):

3d1) subtract |LAB| to each suffix number of the dictionary: $P = P - |LAB|$;

3d2) if the resulting suffix is non-positive, then remove it from the dictionary;

3d3) compute the SA for the encoded LAB $P_{LAB}$ (this sorts the suffixes in the LAB) and update it to: $P_{LAB} = P_{LAB} + |dictionary| - |LAB|$;

3d4) perform a (sorted) insert of these |LAB| suffix numbers into the dictionary.

Fig. 4 illustrates this step for dictionary $D = this\_is\_the\_dict$ and $LAB = \_the$, with lengths 16 and 4, respectively. The LAB is fully encoded and a token is output; then a 4 position SA for the LAB is computed; the value $12 = 16 - 4$ is added to each value in this small SA to get the new indexes of these new suffixes in SA $P$; these new suffixes are inserted into $P$. The contents of $P$ was previously subtracted by 4 in order to update the indexes, according to the actions of slide in and slide out.

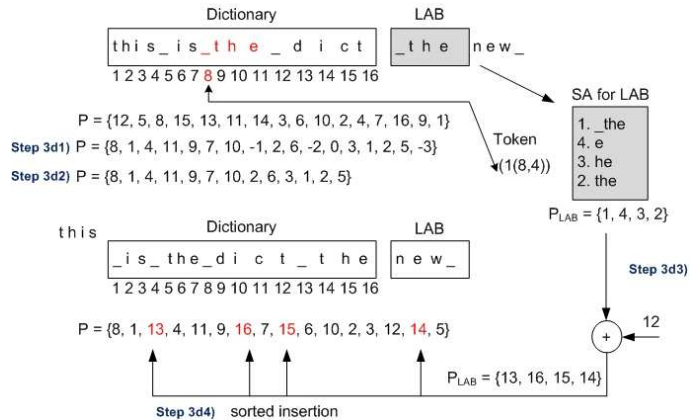

Figure 4: The update step of Algorithm 2 for dictionary $D = this\_is\_the\_dict$ and $LAB = \_the$.

## 3.3 Algorithm 3

The third algorithm uses the concept of LCP as discussed in Section 2.3. The main idea is to **compute the SA and LCP values for the entire sliding window** (the concatenation of dictionary and LAB). This way, we can perform a fast search of the longest sub-strings over the dictionary, that match the longest prefix of the LAB. Fig. 5 depicts the idea of this algorithm, using $D = this\_is\_the\_dict$ and $LAB = \_the$.

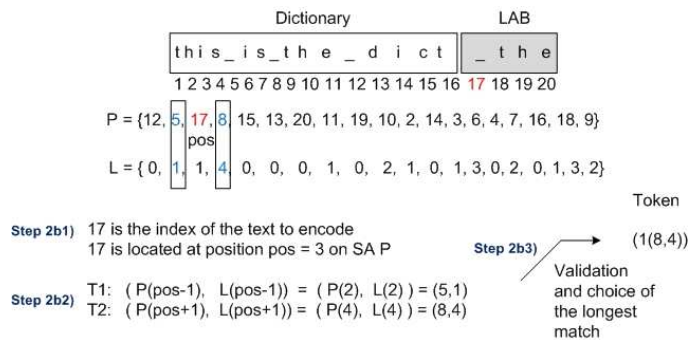

Figure 5: Algorithm 3: use of SA and LCP for LZ77/LZSS encoding.

Notice that, in this case, we have suffixes resulting from the concatenation of the dictionary and the LAB; care has to be taken to encode suffixes of the LAB with position values corresponding to suffixes in the dictionary. This algorithm requires more memory than the previous two, because it uses two integer arrays (SA and LCP) along with the dictionary.

---

**A3 - Algorithm 3 - SA-LCP-Based LZSS Encoding**

Inputs: $In$, input stream to encode; $m$ and $n$ length of dictionary and LAB.
Output: $Out$, output stream with LZSS description of $In$.

---

1. Read dictionary $D$ and look-ahead-buffer $LAB$ from $In$;
2. While there are symbols of $In$ to encode:

    a) Compute SA for $D|LAB$ and name it $P$ and the LCP of $P$, named $L$;
    b) To obtain the description ($pos,len$) for every sub-string $LAB[i \ldots n]$, $1 \leq i \leq n$, proceed as follows:
    b1) Find the position $pos$ of the index $i$ in $P$.
    b2) Use $P$ and $L$ to obtain the two possible tokens $T_1 = (P(pos-1), L(pos-1))$ and $T_2 = (P(pos+1), L(pos+1))$;
    b3) Validate tokens (see if the suffix number is in the dictionary area);
    
    · if both valid, choose and output, to $Out$, the one with the longest match with length $len$; do $i \leftarrow i + len$;
    
    · else if only one token is valid (with length $len$), output it to $Out$; do $i \leftarrow i + len$;
    
    · else (both invalid) output token $(0(LAB[i]))$; do $i \leftarrow i + 1$;
    
    c) Slide in LAB into $D$;
    d) Read next LAB from $In$; goto 2).

---

This algorithm combines the dictionary and the LAB in a single data structure. This way, it is possible to search backward and forward on a given text, making this version suited for the information retrieval scenario.

## 4 Experimental Results

This section presents experimental results of our proof of concept $C$ implementation, using a public SA and LCP package[1]. The tests were carried out using standard files from the well-known Calgary [2] and Canterbury[3] corpora, on a PC with a 2GHz *Intel Core2Duo T7300* CPU and 2GB of RAM. Small and large files from both corpora were used. We report the following results: encoding time and memory, and compression ratio measured in bpb (bits/byte). For comparison purposes, we include the experimental results of an LZSS encoder, based on a binary tree (BT) [15], with 3 integers per tree node; the total number of nodes is |dict.| + |LAB| + 1. The ST-based encoder[4] was written in C++ using Ukkonen's algorithm[4, 5], with an hash table to store each branch (which uses 4 integers) and an array for the nodes (1 integer). The number of branches depends on the contents of the dictionary. Larsson's ST-encoder[5] uses three integers and a symbol for each node, placed in an hash table. For |dict.|=256 and |LAB|=8, Nelson's and Larsson's encoders

---

[1] www.cs.dartmouth.edu/~doug/sarray
[2] links.uwaterloo.ca/calgary.corpus.html
[3] corpus.canterbury.ac.nz
[4] marknelson.us/1996/08/01/suffix-trees
[5] www.larsson.dogma.net/research.html

data structures occupy 3084 and 7440 bytes, respectively. The memory requirements of our algorithms are: |dict.| + |LAB| + |P|, for A1 and A2; |dict.| + |LAB| + |P| + |L|, for A3. In Table 1, we consider an encoding test with a small dictionary (|dict.|=256 and |LAB|=32), using A1, A2, and A3. The amount of memory needed by the SA-based algorithms is 800 = 256 + 32 + 2*256 bytes for A1 and A2, while for A3 we have 1312 = 256 + 32 + 2*256 + 2*256 bytes. For our algorithms, we underline and use bold font for the best encoding time, and use bold font for the second best. There are small differences in the compression ratio, due to the LAB update: the BT-encoder updates the dictionary each time a token is produced; our implementation does this slide in/slide out action, only after a full LAB encoding. The compression ratios could be easily improved by entropy-encoding the tokens.

Table 1: Encoding time (in seconds) and compression ratio (in bpb), for (|dict.|, |LAB|) = (256, 32). The amount of memory is: 800 bytes for A1 and A2; 1312 bytes for A3; 3084 bytes for BT (7440 for ST).

| File | Size | A1 | | A2 | | A3 | | BT | |
|---|---|---|---|---|---|---|---|---|---|
| | | Time | bpb | Time | bpb | Time | bpb | Time | bpb |
| paper5 | 11954 | **0.26** | 5.71 | 0.28 | 5.86 | **<u>0.15</u>** | 5.77 | 0.02 | 5.61 |
| progl | 71646 | **1.03** | 4.28 | 1.15 | 4.51 | **<u>1.00</u>** | 4.49 | 0.07 | 4.15 |
| paper2 | 82199 | **1.20** | 5.82 | 1.36 | 5.94 | **<u>1.11</u>** | 5.86 | 0.08 | 5.89 |
| alice29.txt | 152089 | **2.25** | 5.73 | 2.5 | 5.82 | **<u>2.02</u>** | 5.78 | 0.16 | 5.78 |
| lcet10.txt | 426754 | **5.75** | 5.79 | 6.92 | 5.98 | **<u>2.49</u>** | 5.55 | 0.32 | 5.81 |
| plrabn12.txt | 481861 | **6.95** | 6.11 | 8.37 | 6.18 | **<u>6.08</u>** | 6.14 | 0.48 | 6.16 |

In Table 1, we see that A1 and A2 achieve similar encoding times, and A3 is always faster. If we consider |dict.| = 1024 and |LAB| = 128, we get the results in Table 2, that when compared to Table 1, show lower encoding times and higher compression ratio. From the set of our algorithms, A3 is the fastest (the BT is 3 to 4 times faster), which gives an interesting memory-encoding time trade-off; the update step A2 of is not yet optimized, but it already approaches closely A1 encoding time.

Table 2: Encoding time (in seconds) and compression ratio (in bpb), for (|dict.|, |LAB|) = (1024, 128). The amount of memory is: 3200 bytes for A1 and A2; 5248 bytes for A3; 12300 bytes for BT (29712 for ST).

| File | Size | A1 | | A2 | | A3 | | BT | |
|---|---|---|---|---|---|---|---|---|---|
| | | Time | bpb | Time | bpb | Time | bpb | Time | bpb |
| paper5 | 11954 | 0.14 | 5.51 | **0.11** | 5.73 | **<u>0.06</u>** | 5.68 | 0.02 | 5.18 |
| progl | 71646 | **0.56** | 3.91 | 0.65 | 4.38 | **<u>0.42</u>** | 4.26 | 0.10 | 3.95 |
| paper2 | 82199 | **0.78** | 5.47 | 0.85 | 5.62 | **<u>0.45</u>** | 5.55 | 0.10 | 5.59 |
| alice29.txt | 152089 | **1.37** | 5.41 | 1.57 | 5.54 | **<u>0.86</u>** | 5.49 | 0.22 | 5.39 |
| lcet10.txt | 426754 | **3.46** | 5.45 | 4.56 | 5.66 | **<u>2.46</u>** | 5.55 | 0.37 | 5.38 |
| plrabn12.txt | 481861 | **4.53** | 6.05 | 5.3 | 6.18 | **<u>2.82</u>** | 6.10 | 0.53 | 6.09 |

# 5  Concluding Remarks

We have presented two new SA-based LZ encoding algorithms. These new algorithms improve on earlier work in terms of time-efficient, with similar memory requirements. The key advantage of the new encoding algorithms is that it is possible to *a priori* compute the amount of memory needed to hold the data structures. This may not be the case when using (binary/suffix) trees, because the number of nodes and branches depends on the contents of the text, or when we allocate a memory block that is larger than needed (as it happens with hash tables, to minimize collisions).

Moreover, although not faster than encoders based on binary/suffix trees, our algorithms have considerably lower memory requirements. The low and predictable memory requirement may be important, *e.g.*, in the context of embedded systems, where memory is expensive, and allows the adjustment of the time/memory trade-off. Algorithm 2 is a *sliding window* SA, for which our proof of concept implementation can still be optimized, although the results are quite promising. Algorithm 3 combines the dictionary and the LAB into a single data structure, allowing backward and forward searches; with a large LAB, this algorithm achieves small encoding times. Both algorithms are suited for text search and retrieval, with low and fixed memory requirements, regardless of the contents of the text to search.

In future work, we plan to further optimize Algorithms 2 and 3 in terms of speed, aiming at approaching binary and suffix trees, and explore combinations of Algorithms 2 and 3.

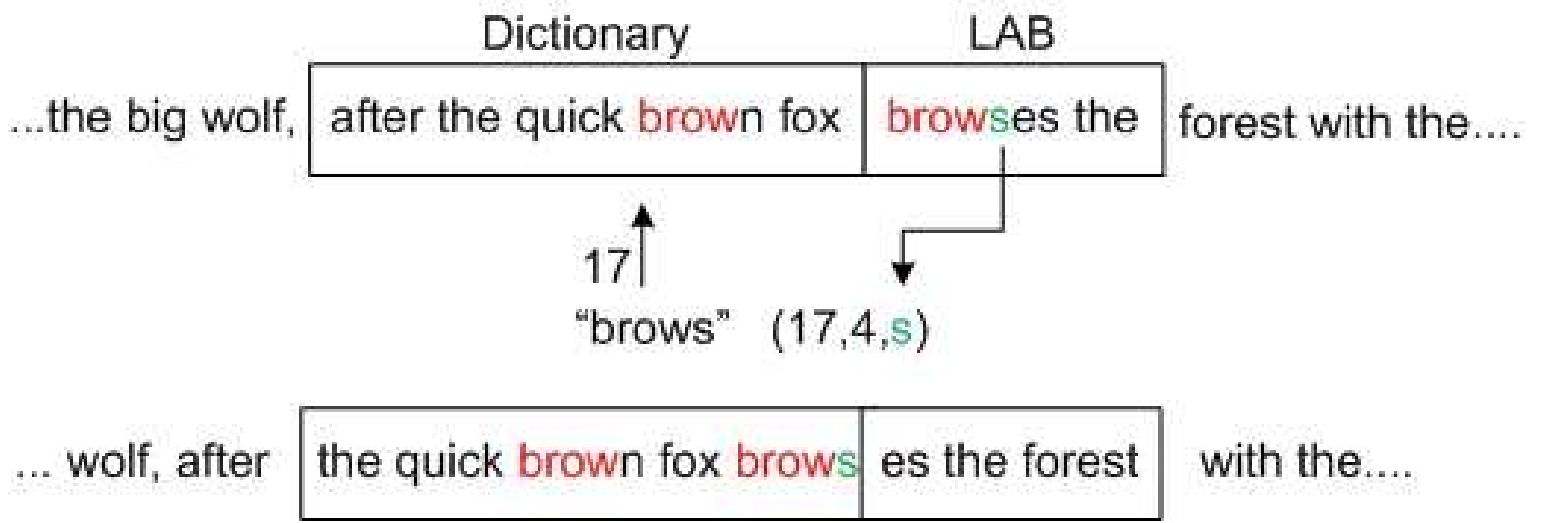

D=mississippi

left=0 → 11. i
    8. ippi
right=3 → 5. 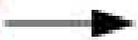ippi
    2. 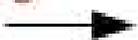issippi
S=issia
    1. mississippi
    10. pi
    9. ppi
    7. sippi
    4. sissippi
    6. ssippi
    3. ssissippi

a)

11. i
8. ippi
5. issippi
2. ississippi
1. mississippi
left=5 → 10. 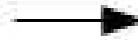pi
right=6 → 9. 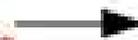ppi
S=psi
7. sippi
4. sissippi
6. ssippi
3. ssissippi

b)